\documentclass[fleqn,usenatbib,useAMS]{aastex631}

\usepackage{amsfonts}
\usepackage{amssymb}

\usepackage{graphicx}
\usepackage{bm}

\usepackage{tikz}
\usepackage{cancel}
\usepackage{braket}
\usepackage{url}
\usepackage{array}
\usepackage{multirow}
\newcolumntype{P}[1]{>{\centering\arraybackslash}m{#1}}
\usepackage{amsmath} 	
\usepackage{amssymb,stackengine,graphicx}

\DeclareMathAlphabet{\pazocal}{OMS}{zplm}{m}{n}

\makeatletter
\@ifundefined{DeclareFontFamilySubstitution}{}{%
  \DeclareFontFamilySubstitution{TS1}{aer}{cmr}}
\makeatother
\IfFileExists{dutchcal.sty}{\usepackage{dutchcal}}{}

\submitjournal{ApJ}

\begin{document}


\title{Comparison of HMG and flat rotation velocities inferred from galaxy-galaxy weak lensing}

\author[0000-0003-3100-2394]{Robert Monjo}
\affiliation{Department of Algebra, Geometry and Topology, Complutense University of Madrid\\ Pza. Ciencias 3, E-28040 Madrid, Spain, rmonjo@ucm.es}
\affiliation{Department of Math and Computer Science, Saint Louis University, Max Aub street, 5, E-28003, Madrid, Spain.}



\begin{abstract}
Despite the success of dark-matter models, unresolved issues require exploring alternatives such as modified gravity theories. In this context, we examine the compatibility of the Hyperconical Modified Gravity (HMG) with galaxy rotation curves inferred from weak-lensing data. The research addresses the existing limitations of Modified Newtonian Dynamics (MOND), which often struggle with universal applicability across different galactic scales. By assuming local validity of General Relativity (GR) and analyzing recent data on circular velocities from galaxy-galaxy weak lensing, our findings interpret the flat rotation curves as an effect of fictitious acceleration inherited from the cosmic expansion, without invoking dark matter. The results indicate that HMG successfully reproduces flat velocity curves on scales of 1 Mpc slightly better than MOND. Therefore, these observations support HMG as a viable gravitational model, highlighting its potential to account for dynamics on galaxies and other scales. Further research with extensive datasets is required to confirm these preliminary insights.
\end{abstract}

\section{Introduction}
\label{sec:intro}

The dark matter (DM) paradigm has been remarkably successful in explaining a wide range of astrophysical phenomena, from the rotation curves of galaxies to the large-scale structure of the universe \citep{Clowe2006, Frenk2012, Bullock2017}. For instance, the Bullet Cluster is often cited as one of the strongest pieces of evidence for DM, as gravitational lensing maps reveal a clear separation between the baryonic mass (hot X-ray gas) and the inferred total mass (from lensing effects), which is difficult to explain in purely baryonic or modified gravity frameworks \citep{Clowe2006}. However, several open issues persist. Despite extensive efforts, DM particles have yet to be detected directly \citep{Roszkowski2018}. Observational phenomena, such as the mass discrepancy-acceleration relation (MDAR) and the baryonic Tully-Fisher relation (BTFR), suggest that our understanding of gravity at galactic scales may be incomplete \citep{McGaugh2016}. In particular, the BTFR is an empirical mass-discrepancy acceleration relation found in galaxy dynamics, which is tightly described as a function of the observed baryonic mass $M_b$ without the need for unobservable matter \citep{Trippe2014, Merritt2017, Goddy2023}. This law is commonly expressed as follows:\begin{eqnarray} \label{eq:BTFR0}
M_b =  A\,v_{\text{flat}}^\epsilon\,,
\end{eqnarray} with $\epsilon \approx 4$ for the level of the \textit{flat} velocity curve $v_{\text{flat}}$ observed in a galaxy disk(see for instance \citealp[][]{Goddy2023}). Similarly, it can be also expressed as a mass-discrepancy acceleration relation (\citealp[][]{McGaugh2004,DiCintio2016}),
\begin{eqnarray} \label{eq:discrepancyMDAR}
\frac{M_{b}+M_{\text{CDM}}}{M_b} =  \frac{v^2}{ v_N^2} = C |a_N|^{-\beta} = C\left(\frac{r}{v_N^2}\right)^\beta\;\; \Longrightarrow\;\; v^{4} = C^2 r^{2\beta}\,v_N^{4(1-\beta)} \approx C^2 \mathrm{G}M_b
\end{eqnarray}where $\beta \approx 0.5$, $M_b+M_{\text{CDM}}$ is the Newtonian dynamical total mass including CDM and $C^2\mathrm{G} \approx: A^{-1}$ is approximately the inverse of the constant used in Eq. \ref{eq:BTFR0}, and the Newtonian acceleration $|a_N| = \mathrm{G}M_b/r = v_N^2/ r$ is expressed as a function of the Kepler-Newton velocity $v_N$ and the radial coordinate $r$ measured from the galaxy center. To explain this phenomenon, some studies suggested that DM presents a stronger coupling to baryons, linking both matter contents by an effective law \citep{Blanchet2007,Katz2016,Barkana2018}. However, most observations suggest the need to modify the standard gravity models because the mass-discrepancy acceleration (with respect the expected Newtonian gravity) only occur where the gravity induced by the visible matter is lower than a typical scale with almost constant value, which points to a more general problem involving scales than a problem involving matter types \citep{Trippe2014, Merritt2017,Comeron2023}. 

Alternative theories such as the Milgromian Dynamics (MOND), Moffat Gravity (MOG) and Hyperconical Modified Gravity (HMG) propose modifications to gravity that could potentially eliminate the need for DM \citep{Milgrom1983, Moffat2009, Milgrom2020, Monjo2023}. 

MOND was introduced by Milgrom in 1983 as an alternative to DM. MOND modifies Newton's second law for accelerations below a certain threshold \(a_0\), leading to the following effective gravitational acceleration: \begin{eqnarray}
\label{eq:mond}
a_{\text{MOND}} = a_{N}\,\nu\left(\frac{a_{N}}{a_0}\right) = \begin{cases}
a_{N} & \text{if } a_{N} \gg a_{0} \\
\sqrt{a_{N} a_0} & \text{if } a_{N} \ll a_{0}
\end{cases},
\end{eqnarray} where $a_0 \approx 1.2 \times 10^{-10} \, \text{m/s}^2$ is the characteristic acceleration scale, while $\nu(x)$ is an interpolation function that satisfies $\nu(x)=1$ for $x \gg 1$ and $\nu(x)=1/\sqrt{x}$ for $x << 1$. The best-fitting $\nu$ function is \citep{McGaugh2016,Banik2022}: \begin{eqnarray}
\label{eq:MOND_interpolation}
    \nu(x)= \frac{1}{1-\exp(-\sqrt{x})}\,.
\end{eqnarray} MOND has been successful in reproducing the flat rotation curves of spiral galaxies without invoking DM \citep{Sanders2003, McGaugh2007}. However, its extension to a relativistic theory still presents some unsolved challenges, in addition to problems at smaller scales \citep{Banik2022, Monjo2023, Banik2023,Cookson2024}.

Hyperconical Modified Gravity (HMG) is a more recent proposal that modifies the gravitational potential by restricting GR to be valid only at a local scale over a background hypeconical metric \citep{Monjo2024}. HMG is a reativistic theory that reproduces the behavior of MOND with a unique natural transition function $\nu$ that addresses the shortcomings of MOND in explaining the radial acceleration relation (RAR) of galaxy clusters \citep{MB2024}. Specifically, the HMG transition function depends on the distribution of mass and the geometry of spacetime.

To complete our previous findings, this \textit{letter} analyzes the HMG modeling for flat rotation curves, inferred from weak lensing, recently published \citep{Mistele2024}. To analyze the theoretical circular orbital velocities from weak lensing using a specific Lorentzian metric, the authors widely considered the case of a spherically symmetric metric given by $ds^2 = f(r)c^2 dt^2 - g(r) dr^2 - r^2 (d\theta^2 + \sin^2\theta \, d\phi^2)$, where $f(r)$ and $g(r)$ are functions of the radial coordinate $r$ \citep{Bartelmann2001, bradac2005}. The enclosed mass $M(r)$ within a radius $r$ is related to the metric functions $f(r)$ and $g(r)$. For a static and spherically symmetric spacetime, $f(r) = g(r)^{-1} =  1 - 2{\Phi}/{c^2}$ imitates the Scwarzschild metric, with gravitational potential at $r$ defined by $\Phi(\mathbf{r}) \equiv -G \int {\rho(\mathbf{r'})}/{|\mathbf{r} - \mathbf{r'}|} \, dV'$, where $\rho(r')$ is the mass density at a point $r'$ distributed around the volume $V'$, and approaches $\Phi(r) = {GM(r)}/{r}$ in the classic spherical limit. In the Newtonian regimen, with weak gravity ($r \gg 2GM/c^2$) and low velocities ($v/c \ll 1$), the function $f(r)$ contributes significantly more than the function $g(r)$ to geodesics \citep{Bartelmann2001, dolan2023}. 


As a key in weak lensing analysis, the surface mass density $\Sigma(R)$ at projected distance $R$ is obtained by projecting the mass density $\rho(r)$ along the line of sight according to $\Sigma(R) = 2 \int_{0}^{\infty} \rho(\sqrt{R^2 + z^2})\, dz$, and the \textit{convergence} or dimensionless surface mass density is given by $\kappa(R) = {\Sigma(R)}/{\Sigma_{\text{crit}}}$, where $\Sigma_{\text{crit}} = \frac{c^2}{4\pi G} {D_s}D_d^{-1} D_{ds}^{-1}$ is the critical surface mass density for lensing, while $D_s$, $D_d$, and $D_{ds}$ are the angular diameter distances from the observer to the source, the observer to the lens, and the lens to the source, respectively \citep[see figure 1 of][]{Banik2015}. These relationships allow us to derive the mass profile from radial acceleration $a_\text{obs}(r)$ and subsequently calculate the theoretical circular orbital (square) velocity $v_c^2(r) := r\, a_\text{obs}(r)$, which is critical for understanding the dynamics of galaxies through weak lensing data \citep{bradac2005, Brouwer2021, Mistele2024}.

\section{Data and model}
\label{sec:obs_and_model}
\subsection{Inferred observations}

This \textit{letter} uses the published results of \cite{Mistele2024} on the observed circular velocity inferred from weak leansing and classified in four bins of baryonic mass (1.29, 4.57, 9.13, and 19.5 in units of $10^{10} M_\odot$). Specifically, we use 20 accurate estimates of the \textit{flat} speed ($v_{\text{flat}}$) as a function of the observed mass and distances (table 1 of \citealp[][]{Mistele2024}). Moreover, our work uses baryonic Tully-Fisher data published in the same paper, distinguishing between ranges of 300 kpc and 1 Mpc, as well as between early-type galaxies (ETGs) and late-type galaxies (LTGs). The results are consistent with \cite{Lelli2019}.

\subsection{Metric perturbation of HMG}

Observed orbital speed data inferred from weak lensing were compared to the values modeled by HMG, for the observed baryonic mass, according to the work developed by \citet{Monjo2023} and \citet{MB2024}, which is summarized here. Let $g$ be the background metric of the hyperconical universe \citep{Monjo2017,MCS2023}. The metric $g$ is locally approximately given by
\begin{eqnarray} \label{eq:hyp1}
g \approx dt^2 \left(1- kr'^2 \right) 
-  \frac{t^2}{t_{0}^2} \left( \frac{dr'^2}{1-kr'^2} + {r'}^2d{\Sigma}^2 \right)
-  \frac{ 2r't}{t_{0}^2} {dr'dt}\,,
\end{eqnarray}where $k = 1/t_0^2$ is the spatial curvature for the current value of the age $t$ of the universe ($t_0 \equiv 1$), while $t/{t_0}$ is a linear scale factor, $r' \ll t_0$ is the comoving distance, and $\Sigma$ represents the angular coordinates. The shift and lapse terms of Eq. \ref{eq:hyp1}, produced by the comoving observers, lead in torn to an apparent radial spatial inhomogeneity that is assimilated as a fictitious acceleration with adequate projected coordinates.

The HMG model is based on two key ideas: \begin{enumerate}
\item Any gravitational system of mass $M_{\text{sys}}$ generates a perturbation over the diagonal terms of the background metric $g \to \hat{g}$ in Eq. \ref{eq:hyp1}, such that $k{r'^2} \to k{{r}'^2} + \int_{r'}^\infty {2GM(r)}/(c^2 r^2) dr$ with $t \approx t_0$. In other words, GR is only (locally) valid with respect to the background metric \citep{Monjo2024}, thus gravity dynamics is due to the perturbation term $\hat{h} := \hat{g}-g$,\begin{eqnarray}  \label{eq:Schwarzschild_g}
  g_{Sch}  :\approx \eta + (\hat g - g) \approx  \left(1- \int_{r'}^\infty \frac{2GM(r)}{c^2 r^2} dr \right) c^2 dt^2
-  \left[\left(1 + \int_{r'}^\infty \frac{2GM(r)}{c^2 r^2} dr\right) dr'^2 + r'^2d{\Sigma}^2 \right]\,;\;\;\;\;\;\;
\end{eqnarray}
\item the coordinates that parameterize the metric, are finally projected as follows:\begin{eqnarray} \nonumber
    r' \;\; &\to& \;\; {\hat r}' = \lambda^{1/2}r' 
    \\ \nonumber
    t \;\; &\to& \;\; \hat t = \lambda t
\end{eqnarray} where $\lambda := 1/(1-\gamma/\gamma_0)$ is the stereographic scaling, which is a function of the angular position $\gamma = \sin^{-1}(r'/t_0)$ and a projection factor $\gamma_0^{-1} = \gamma_{\text{sys}}^{-1}\cos \gamma_{\text{sys}}$, where $\gamma_{\text{sys}}$ is the characteristic angle of the gravitational system. In an empty universe, $\gamma_0 = \gamma_U / \cos \gamma_U$. We expect $\gamma_U = \frac{1}{3}\mathrm{\pi}$ and therefore that $\gamma_0 = \frac{2}{3}\mathrm{\pi} \approx 2$ \citep{MCS2023}. \end{enumerate}

With these ingredients, minimally-perturbed metric derived by HMG outside a distribution mass is (Appendix \ref{sec:circular}):
\begin{eqnarray} \label{eq:deflection_HMG0}
\hat g &\approx& \left(1 - \frac{2GM}{c^2 r} + \frac{2r}{\gamma_0(r) c t}\right) c^2 dt^2  - \left[\left(1 + \frac{2GM}{c^2 r} + \frac{2r}{\gamma_0(r) c t}\right) dr^2 + r^2 \left(1 + \frac{r}{\gamma_0(r) c t}\right) d\Sigma^2\right]\,,
\end{eqnarray}where $r/\gamma_0(r)$ approaches to 0 for both $r \to 0$ and $r \to \infty$. Alternatively, comoving radial coordinate $r' \equiv (t_0/t)\, r$ can be used instead of $r$, but it is assumed to be approximately equivalent for gravitational systems close to the observers.


The spacetime derived from local GR (that is, resulting in Eq. \ref{eq:Schwarzschild_g}) and the one represented by the perturbed metric (Eq. \ref{eq:deflection_HMG0}) satisfy the requirement to be asymptotically flat and coincide with the diagonal Schwarzschild solution at large distances. Therefore the effective ``enclosed mass'' can be estimated as usual in weak lensing \citep{Bartelmann2001, will2014, umetsu2020}. To illustrate the behavior of Eq. \ref{eq:deflection_HMG0} for weak gravitational fields, an example of deflection angles is derived in Appendix \ref{sec:deflection_HMG}. 

\subsection{Orbital speed and Hubble flux relation}

The orbital speed $v_{\text{C}}$ is modeled by HMG according to the geodesic equations obtained for the projected time component of the perturbation $\hat{h}_{tt}$. As a result, a fictitious cosmic acceleration of $a_{\gamma 0}(r) := \gamma_0(r)^{-1}2c/t = \gamma_{\text{sys}}^{-1}(r)\cos\gamma_{\text{sys}}(r)\, 2c/t$ emerges with a non-negligible time-like component, which contributes to a more effective gravity with a total centrifugal acceleration $a_{C}(r) = v_{\text{C}}^2/r$ such that (Appendix \ref{sec:circular}):\begin{eqnarray}
\label{eq:centrifugal}
    a_{C}(r)^2 \approx  a_N(r)^2 + |a_N(r)|\frac{2c}{\gamma_0(r) t} \;\;\Longrightarrow\;\; \left(\frac{v_{C}}{ v_N}\right)^2 \approx \sqrt{1 + \frac{1}{|a_N|} \frac{2c}{\gamma_0(r) t}} \,,
\end{eqnarray} where $a_N(r) := {G}M_{\text{sys}}/r^2$ is the Newtonian acceleration and $v_{N} = \sqrt{a_N(r)r}$ is the circular orbital speed, while the last term (that depends on $\gamma_0$) represents an additional contribution to the gravitational acceleration. Notice that Eq.~\ref{eq:centrifugal} leads to $v_{\text{flat}} = \lim_{r\to\infty} v_{C} = [2{G}M_{\text{sys}}c/(\gamma_0 t)]^{1/4}$ as in Eq.~\ref{eq:BTFR0}. However, assuming that gravitationally-bounded space does not expand (i.e., $r'(t_0) \mapsto r(t) =r'(t_0)$ does not increase) as a function of time $t_0 \mapsto t>t_0$, one could require to use the redshift $z$ for estimating the observed time $t \mapsto t_0 \approx t/(1+z/2)$ in the gravitational system \citep{Monjo2024}. 

For modeling observations of circular orbital velocities, the characteristic angle $\gamma_{\text{sys}}$ of the considered system can be estimated from a general approach by considering the relative geometry (angle) between the Hubble speed $v_H(r) := r/t$ and the Newtonian circular speed $v_{N}(r) := \sqrt{{G}M_{\text{sys}}/r}$, as follows \citep{MB2024}: \begin{eqnarray}
\label{eq:model_general2}
\sin^2 \gamma_{\text{sys}}(r) \;\;\; & \approx & \;\;\;  \sin^2\gamma_U + \left({\sin^2\gamma_{\text{center}} -  \sin^2 \gamma_U} \right) \bigg| \frac{2v_N^2(r)-\epsilon^2_H v_H^2(r)}{2v_N^2(r) + \epsilon^2_H v_H^2(r)}\bigg|\;,
\end{eqnarray}where the parameter 
$\epsilon^2_H$ is the so-called \textit{relative density of the neighborhood}, while $\gamma_{\text{center}} \approx \mathrm{\pi}/2$ and $\gamma_U = \mathrm{\pi}/3$ can be fixed here to set a 1-parameter ($\epsilon_H$) general model from Eq. \ref{eq:model_general2} instead of fitting an average value $\gamma_0$ in the simple $\gamma$-HMG version (Eq. \ref{eq:centrifugal}). 

The interpretation of Eq. \ref{eq:model_general2} is the following: The additional contribution to the gravity in galaxy rotation curves is minimal when the projective angle is maximal (i.e., $\gamma_{\text{sys}} = \gamma_{\text{center}}$), reached for the highest velocities $v_{N} \gg \epsilon_Hv_H$ at small radial distances (i.e., recovering the classical Newtonian regimen). In contrast, the additional contribution is maximum with minimum angle, $\gamma_{\text{sys}} = \gamma_U$, when the Newtonian speed is reduced down to the Hubble flux $v_{N} = \epsilon_Hv_H$ at larger distances (bordering with neighbor systems). Therefore, the model described by Eq.~\ref{eq:model_general2} consists of a key parameter $\epsilon_H$ that refers to the dynamical equilibrium of the neighborhood (that is, $v_{N} = \epsilon_Hv_H$)and modulates the more effective gravity.

\begin{figure}[h]
\centering
	\includegraphics[width=0.75\columnwidth]{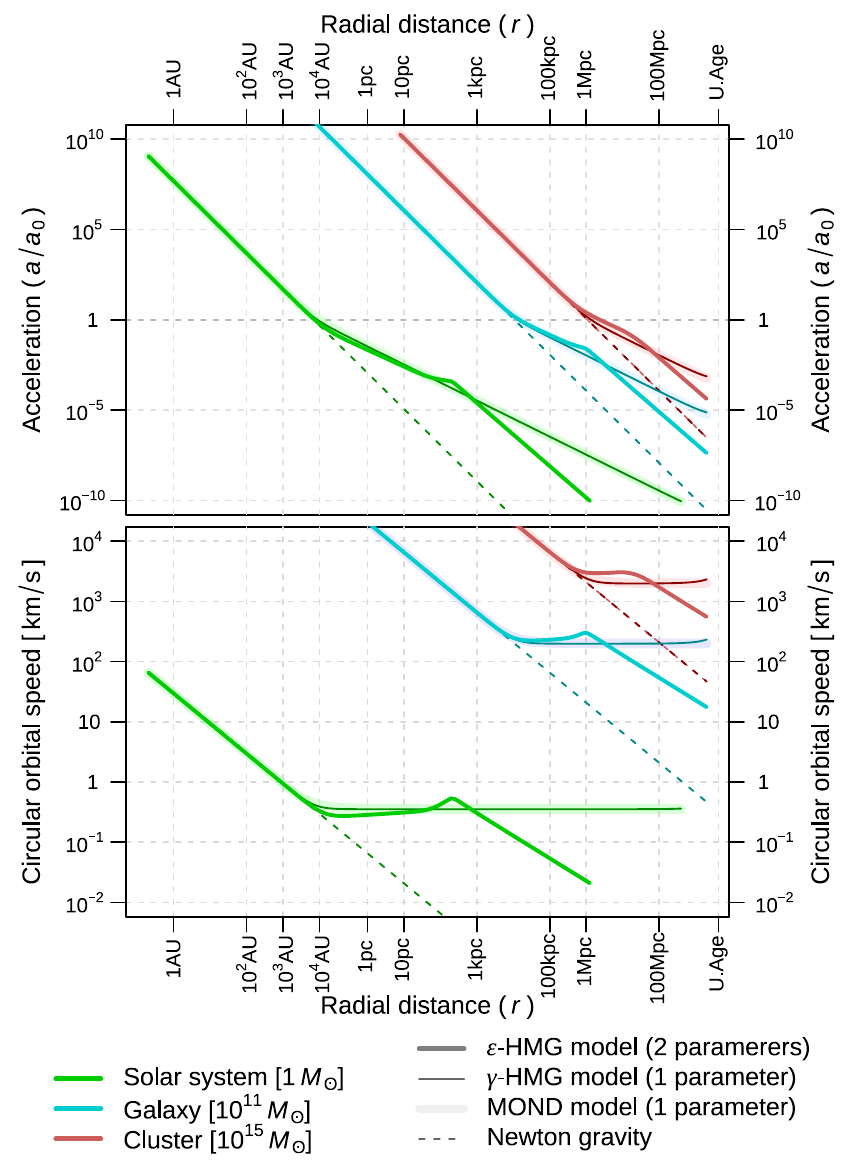}
    \caption{Example of acceleration (top panel) and circular orbital speed (bottom panel) curves, as a function of the radial distance ($r$), for three gravitational systems centered, respectively, by three characteristic baryonic masses enclosed in $r$: Sun ($1 M_\odot$; green), galaxy ($10^{11} M_\odot$, blue) and galaxy cluster ($10^{15} M_\odot$, red). The models used are: the classical Newtonian gravity (dashed lines), an empirical MOND function (Eq. \ref{eq:MOND_interpolation} with standard $a_0 = 1.2 \times 10^{-10} \text{m/s}^2$; lightly colored width lines), $\gamma$-HMG (Eq. \ref{eq:centrifugal} with $\gamma_0 = 12$; thin colored lines) and $\epsilon$-HMG (Eqs. \ref{eq:model_general2} and \ref{eq:sigma_model} with $r_\text{sys} \equiv r$ and different values of $r_\text{nei}/r_\text{sys}$ and $\zeta$; width colored lines). The represented $\epsilon$-HMG configurations ($r_\text{nei}/r_\text{sys}$, $\zeta$) are: (4, 0.9) for the Solar system, (3, 0.9) for the galaxy, and (2, 0.9) for the cluster. The value of $\zeta \equiv 1$ implies that a gravitational system is close to equilibrium with respect to the Hubble flux. The radial distances ($r$) range between 1 Astronomic Unit (AU) and the maximum \textit{light-travel distance} ($t_0c$) given by the age $t$ of the Universe (noted as `U. Age', which is $t_0c \approx 13.7\,\text{Gyr}\,c$ with $c \equiv 1$).}  \label{fig:Fig00}
\end{figure}

In other words, the parameter $\epsilon_H$ is not totally free, since it is theoretically and statistically strongly linked to the observed density $\rho(r)$ of the system at $r = r_{\text{nei}} \sim 50-200\,$kpc, which are neighborhood distances $r_{\text{nei}}$ with dynamical equilibrium \citep{MB2024}. In particular, for a typical distance $r_{\text{sys}} < r_{\text{nei}}$, we found that $\frac{1}{6} \le \epsilon_{H}^2(r_{\text{sys}}) \le \frac{1}{6}+ {\rho(r_{\text{nei}})}/{\rho_{\text{vac}}}$, where $\rho_{\text{vac}} := 3/(8\pi G t^2)$ is the vacuum density. To test its impact on the model, we considered two approaches in this work. On the one hand, we set $\gamma_0(r) = \gamma_0$ as a characteristic constant (at least for each type of system) to simply fit an 1-parameter $\gamma$-HMG model (Eq. \ref{eq:centrifugal}) in a similar way as in the MOND paradigmn. On the other hand, we assume an $\epsilon$-HMG version in which $\gamma_0(r) = \gamma_{\text{sys}}(r)/\cos\gamma_{\text{sys}}(r)$ depends on the velocities modulated by $\epsilon_H^2$ as shown in Eq. \ref{eq:model_general2}. In torn, $\epsilon_H$ is presumed to vary as a function of the system density $\rho(r_{\text{sys}})$ at a given distance $r_{\text{sys}}$, as follows:\begin{eqnarray}
\label{eq:sigma_model}
   \epsilon_H^2(r_{\text{sys}}) - \frac{1}{6} \; \approx \; \left(\frac{\rho(r_{\text{nei}})}{\rho_{\text{vac}}} \right)^\zeta = \left(\frac{\rho(r_{\text{sys}})}{\rho_{\text{vac}}}\right)^\zeta \left(\frac{r_{\text{sys}}}{r_{\text{nei}}}\right)^{3\zeta}\,, 
\end{eqnarray}where $\zeta = 1$ is theoretically derived but a range is empirically found about $0.8 < \zeta \le 1$ \citep{MB2024}. Therefore, the semi-empirical $\epsilon$-HMG has two parameters, $\zeta$ and $r_{\text{nei}}/r_{\text{sys}} \ge 1$, which modulate the equilibrium of the gravitational system (Fig. \ref{fig:Fig00}).

Notice that when $\zeta = 1$ and $r = r_{\text{sys}}$ with $\epsilon_H \gg \frac{1}{6}$, the $\epsilon$-HMG model is equivalent to the simplest $\gamma$-HMG version, for which the Hubble-Newton dynamical equilibrium ($\gamma_{\text{sys}} = \gamma_{U}$) is reached at $r = r_{\text{sys}} = r_{\text{nei}}$ in Eq. \ref{eq:model_general2}, as it was found for some galaxies in the same study. In contrast, if $\zeta < 1$, the system is not in equilibrium for $r=r_{\text{sys}}$ under the $\epsilon$-HMG approach (Eqs. \ref{eq:model_general2} and \ref{eq:sigma_model}). Thus, we set $\zeta \approx 0.8$ to estimate $r_{\text{nei}}/r_{\text{sys}}$ for galaxies whose circular orbital speed is clearly higher than the Hubble flux. Finally, flat velocities were modeled by fitting $a_0$ in MOND and $\gamma_{\text{sys}}$ in $\gamma$-HMG, since these constant speeds are independent of radial distances due to their own asymptotic definition (i.e., $\epsilon$-HMG cannot be adequately fitted). Of the 30 rows of Table 2 of \citet{Mistele2024}, only 22 are statistically independent, namely the 8 LTG, 6 ETG and 8 kinematic entries; the remaining 8 rows are aggregates of the LTG and ETG samples and repeat the same galaxies, so they are shown in the figures but excluded from every $\chi^2$ sum, which therefore has 21 degrees of freedom.

\section{Results and discussion}
\label{sec:results}


\begin{table}[h]
    \centering
     \caption{\label{tab:1}Fitting parameters of an empirical MOND interpolation (Eq. \ref{eq:MOND_interpolation}), $\gamma$-HMG (Eq. \ref{eq:centrifugal} with $\gamma_0 = \text{constant}$) and $\epsilon$-HMG (Eqs. \ref{eq:model_general2} and \ref{eq:sigma_model} with $\zeta=0.8$) models to observed circular and flat velocities inferred from weak lensing \citep{Mistele2024}.}
    \label{tab:my_label}
    \begin{tabular}{ccccc}
   \hline
    \hline
    Method & Velocity data & Parameter & $\chi^2$ (p-value) & Best one (*)  \\
		\hline
MOND & circular & \;\;\;\;$a_0 = 1.21_{-0.08}^{+0.09} \times 10^{-10} m/s^2$ \;\;\;\; & 31.3 (0.96) & - 
\\
$\gamma$-HMG & circular &  $\gamma_0 = 9.8_{-0.6}^{+0.7}$ & \;\; 25.7 (0.86) \;\; & -
\\	
\;\;\;\;$\gamma$-HMG + z\;\;\;\; & circular &  $\gamma_0 = 11.9_{-0.8}^{+0.9}$ & 25.9 (0.87) & -
\\	
$\epsilon$-HMG & circular &  $r_{\text{nei}}/r_{\text{sys}} = 2.54_{-0.12}^{+0.14}$ & 18.0 (0.48) & - 
\\
$\epsilon$-HMG + z & circular & $r_{\text{nei}}/r_{\text{sys}} = 2.95_{-0.12}^{+0.18}$ & 17.5 (0.44) & * 
\\
    \hline
MOND & flat & $a_0 = 1.60_{-0.05}^{+0.06} \times 10^{-10} m/s^2$ & 23.2 (0.67) & - 
\\
$\gamma$-HMG & flat &  $\gamma_0 = 8.3_{-0.3}^{+0.3}$ & 22.0 (0.60) & *
\\	
$\gamma$-HMG + z & flat &  $\gamma_0 = 9.7_{-0.3}^{+0.4}$ & 22.3 (0.62) & -
\\
    \hline
    \end{tabular}

    \vspace{2pt}
    \begin{minipage}{\columnwidth}
    \small \textsc{Note}: The symbol $z$ is added when redshift is considered via $t_0 \approx t/(1+z/2)$ in the fitting of $\gamma$-HMG and $\epsilon$-HMG. Uncertainties are the $1\sigma$ profile intervals of the grid scan, that is, the range of parameter values with $\chi^2 \le \chi^2_{\text{min}} + 1$. The $\chi^2$ of the flat-velocity fits is summed over the 22 statistically independent rows of Table 2 of \citet{Mistele2024} (see Sect. \ref{sec:obs_and_model}).
    \end{minipage}
\end{table}

\begin{figure}[h]
\centering
	\includegraphics[width=1.01\columnwidth]{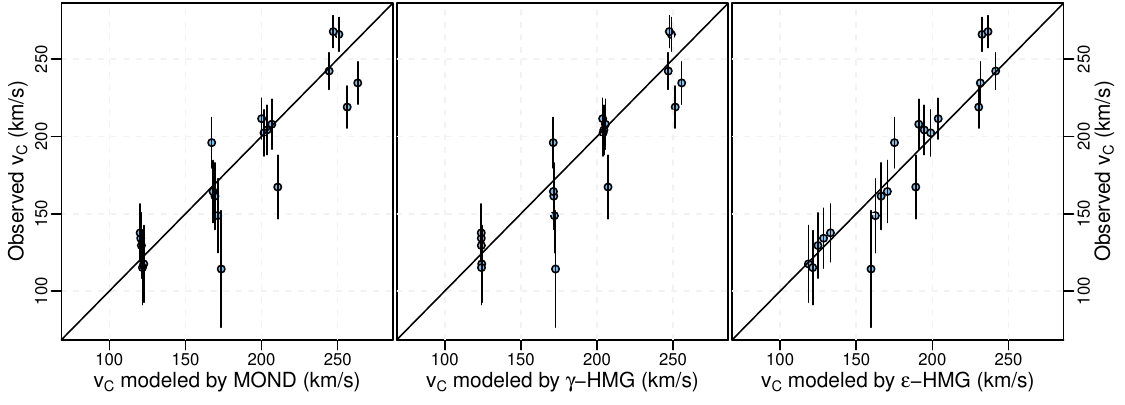}
    \caption{Modeling of circular speed assuming the current time $t\approx t_0$ (i.e., no redshift effects) according to an empirical MOND function (Eq. \ref{eq:MOND_interpolation} with $a_0 = 1.21 \times 10^{-10} \text{m/s}^2$; left panel), $\gamma$-HMG (Eq. \ref{eq:centrifugal} with $\gamma_0 = 9.8$; medium panel) and $\epsilon$-HMG (Eqs. \ref{eq:model_general2} and \ref{eq:sigma_model} with $\zeta = 0.8$ and $r_\text{nei}/r_{sys} = 2.5$; right panel). These models are fitted to the observed circular velocity inferred from weak leansing (Table 1 of \citealp[][]{Mistele2024}). Data are gathered in four sets according to four baryonic mass bins (1.29, 4.57, 9.13, and 19.5 in units of $10^{10} M_\odot$).\\
    {}
}  \label{fig:Fig01}
\end{figure}

\begin{figure}[h]
\centering
	\includegraphics[width=0.82\columnwidth]{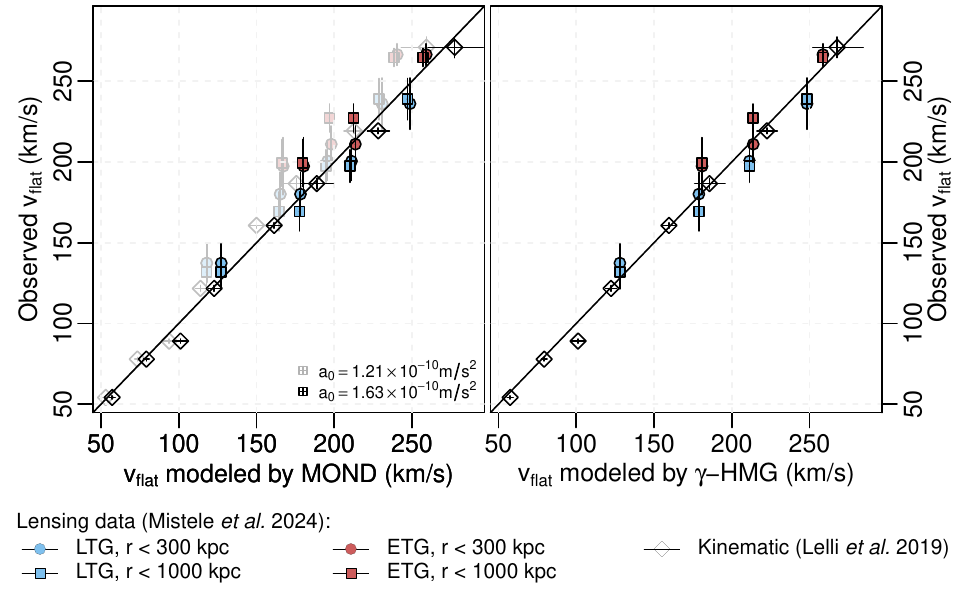}
    \caption{Flat velocity (asymptotic speed of the rotation curve) according to weak lensing observations for ETGs and LTGs separately (red and blue symbols; \citealp[][]{Mistele2024}) and kinetic data (white diamonds; \citealp[][]{Lelli2019}). The left panel shows values from the empirical MOND function with $a_0 = 1.21 \times 10^{-10}\text{m/s}^2$ (light colors), fitted with circular speeds (Table \ref{tab:1}) and with $a_0 = 1.60 \times 10^{-10}\text{m/s}^2$ (dark colors) fitted to flat velocities. The right panel shows the $\gamma$-HMG model fit to flat velocities with $\gamma_0 = 8.3$.\\
    {}
}  \label{fig:Fig02}
\end{figure}

\subsection{General findings}

The $\gamma$-HMG and $\epsilon$-HMG versions of the model and their derived dynamics illustrate the motion of circular orbits in a modified spacetime, reflecting both classical Newtonian and non-classical gravitational effects due to a distorting acceleration of about ${c}/(\gamma_0 t) \lesssim 10^{-10}\,\text{m/s}^2$. The resulting new terms represent cosmic influences on gravitational phenomena that should be observable in large-scale astrophysical contexts, as our results suggest.

Without considering redshift effects, the best $\gamma$-HMG model fitted to \textit{flat} velocities is found for $\gamma_0 = 8.3_{-0.3}^{+0.3}$, which lies $2.1\sigma$ below the lowest value of $\gamma_0 = 9.8_{-0.6}^{+0.7}$ fitted to circular velocities (Table \ref{tab:1}). This ordering is not coincidental because $\gamma_0$ is not a constant in the HMG paradigm since it decreases for galaxies (Eq. \ref{eq:model_general2}) as the radius increases with an almost flat velocity. The same ordering is found when the redshift effect is considered: the lowest value of the projection factor for circular velocities ($\gamma_0 = 11.9_{-0.8}^{+0.9}$) again exceeds the one obtained for \textit{flat} velocities ($\gamma_0 = 9.7_{-0.3}^{+0.4}$), here by $2.5\sigma$.

On the other hand, the MOND parameter $a_0 = 1.21_{-0.08}^{+0.09} \times 10^{-10} m/s^2$ fitted to circular velocities is incompatible with the \textit{flat} velocities analyzed in this work (Fig. \ref{fig:Fig02}), and is $3.7\sigma$-separated with respect to the best fit ($a_0 = 1.60_{-0.05}^{+0.06} \times 10^{-10} m/s^2$). Both scales are derived from the same lensing masses, so a common offset in the stellar mass-to-light ratio largely cancels in their difference. It does not cancel when either of them is compared with the standard value of $a_0 = 1.2 \times 10^{-10} m/s^2$, which is assumed to be a universal constant: the flat-velocity scale exceeds it by more than seven times the statistical uncertainty of the fit, but that interval carries no mass-to-light systematic and this figure should therefore be read as an upper bound. Among the eight fits of Table \ref{tab:1}, the circular-velocity MOND fit is the poorest description of its own data ($\chi^2$ percentile of 0.96). Moreover, the fitting of the $\epsilon$-HMG model showed qualitatively better results in explaining the circular velocities self-consistently compared to the MOND and $\gamma$-HMG models (Fig. \ref{fig:Fig01}). 

The impact of the redshift on the above findings is not statistically significant for modeling observations of circular orbital velocities (Table \ref{tab:1} and Fig. \ref{fig:Fig01}). The explanation is that gravity strongly binds its spatial coordinates, mostly removing the effect of expansion, so $r = r'$. In contrast, the flat velocity $v_{\text{flat}}$ approaches the external limit of the gravitational force, and therefore the Hubble law $r'/t_0 \approx r/t$ is approximately satisfied. From this effect, the observed cosmic acceleration is practically invariant under the observable time periods (i.e., the dependence on redshift is weak). This finding is consistent with emergent data on the rotation curves of disk galaxies at significant redshifts \citep{Bekenstein2008, McGaugh2016}. 

\vspace{5mm}

\subsection{Limitations of the current HMG model}

The standard $\Lambda$CDM model has achieved numerous successes such as the ability for reproducing the perturbation spectrum in the cosmic microwave background (CMB), the observed large-scale structure of the Universe, and the formation of galactic structures among other astrophysical observations \citep{Dodelson2003,Li2020}. It is well established that purely baryonic disk galaxies, in the absence of DM halos, are highly susceptible to bar instabilities and excessive spiral structure growth. Numerical simulations and theoretical studies have consistently shown that disk galaxies without a stabilizing mass component tend to develop bars and rapidly evolve away from their observed morphologies \citep{Ostriker1973,Efstathiou1982,Weinberg2002}.

Any modified gravity should count whether it can stabilize the disks against bar formation. \citet{Tiret2007} explored the performance of MOND in structure formation and found that galactic discs experience bar instability earlier than the DM model. Additionally, the evolution of bar pattern speed differs significantly because MOND systems experience significantly weaker dynamical friction compared to equivalent Newtonian systems, (i.e., with the same phase-space distribution of baryons and dark matter; \citealp[][]{Nipoti2008}).

Since HMG recovers the MOND regimen for most of the observed galaxy dynamics, it is probably challenged to stabilize the formation of galactic systems. Therefore, it is necessary to analyze how HMG may (or may not) offer a natural stabilization mechanism without the need for DM halos. The mechanism of \textit{fictitious acceleration} (that emerges from the cosmic background metric) could be a key to stabilization.

Another critical point is that systems of comparable mass exhibit distinct dynamical properties, which result in different needs (or no need) for DM components. For example, ultra-faint galaxies, within the framework of Newtonian gravity, are among the most DM-dominated systems, exhibiting exceptionally high inferred DM-to-baryon ratios—often exceeding those estimated for galaxy clusters \citep{Simon2019}. In contrast, some globular clusters with a mass range similar to that of many ultra-faint galaxies do not appear to require a significant DM component \citep{Moore1996,Forbes2017}. One notable example is the galaxy NGC 1052–DF2, which has been observed to have an unusually low dark matter content \citep{Forbes2017,Baumgardt2018}.  This striking diversity raises a fundamental question that HMG should address in future work. A possible way to solve this issue could be the proposal of \citet{MB2024}, who explored the ability of HMG for reproducing a multiscale range of distances, accelerations, and orbital velocities.

However, the standard DM-based model also present difficulties to explain some open issues. For example, numerical simulations predict that DM halos should have central cusps, while those inferred from observed galaxies do not have cusps \citep{Weinberg2002}. Moreover, the observed number of small galaxies and dwarf satellite systems in the Local Group is significantly lower than the predicted abundance of low-mass dark matter halos and subhalos within comparable volumes. These discrepancies contribute to some of the most well-known challenges faced by the $\Lambda$CDM paradigm, including the cusp/core problem, the missing satellites problem, and the too-big-to-fail issue \citep{Bullock2017}. The lack of direct detection of DM and the problems in modeling structures such as the \textit{impossible early galaxies} \citep{Yennapureddy2018}, the \textit{Giant Arc} and \textit{Big Ring} \citep{Lopez2024} or the \textit{El Gordo} collision \citep{Asencio_2023} could promote the development of alternatives such as the HMG model or other MOND-like theories  \citep{Kroupa2015,Asencio2022,kroupa2023}. For example, the Hubble tension on the different acceleration rates of the Universe could be addressed by the fictitious acceleration that emerges from the \textit{hyperconical model}, which is used as a basis for the HMG \citep{Monjo2018,Monjo2024}.

Hence, it is essential to investigate whether the HMG model naturally leads to solving some of the above problems, including a wide range of multi-scale structure formation, with adequate stability.

\section{Concluding remarks}
\label{sec:conclusions}

According to the HMG model, the Newtonian gravitational acceleration is modified by a distorting term $c/(\gamma_0 t)$ that represents a fictitious acceleration inherited from the large-scale effects of comoving reference frames. The modified gravity results in an additional multiscale cosmic acceleration offset but with a special print on galactic dynamics. It worth recalling that gravitational lensing is derived from deviations in the metric geodesics, with a deeper curvature that would be caused by dark matter or by a more efficient gravity from baryonic matter. 

The analysis of circular velocities inferred from weak lensing observations suggests the viability of HMG as an alternative to the \textit{dark matter} paradigm, providing a relativistic cosmic approach to the MOND regime. By fitting the observed data to the model, we demonstrate that HMG can reproduce the flat velocity curves even for the 1 Mpc scale, without requiring unobservable particles. Our findings showed that the circular and flat rotation curves are statistically independent of the readshift, which suggests that the coupling of the cosmic acceleration to the perturbed metric in the local GR is very weak. However, the wide-scale flexibility of fitting and the cosmological basis of HMG provide a consistent and potentially more fundamental explanation for the observed gravitational phenomena in galaxies. Although our preliminary findings reinforce the viability of the theory, future studies with larger datasets and more refined versions are needed to validate these results and explore the full implications of HMG on our understanding of cosmology and gravity. Complementary work is now being carried out aligned with reproducing the cosmic microwave background spectrum and the Big Bang nucleosynthesis using the timeline simulated by \cite{MCS2023}. In future works, we will explore whether effective modifications of the gravitational potential induced by the fictitious acceleration of HMG could contribute to disk stability. Moreover, we will examine the ability of HMG to predict the correct behavior of merging clusters and to reproduce observations such as the \textit{Bullet Cluster} or \textit{El Gordo} collisions, among other large-scale structures.

 \smallskip
  \smallskip

\section*{Acknowledgements}

The authors thank T. Mistele for providing all the necessary data for this study, as well as for all comments and suggestions. We also acknowledge the contributions of our colleague I. Banik for the valuable feedback and discussions, especially on the interpretation of multiscale behavior of the HMG model. Finally, we greatly appreciate the very constructive comments of the reviewers that have helped to improve this article. We are also indebted to V. \v{C}e\v{c}elev for an independent reproducibility audit that identified the non-independence of the aggregate rows of the lensing sample and prompted the revision of the uncertainty conventions reported in Table \ref{tab:1}.

 \smallskip
 
\section*{Data Availability}

In this study, no new data was created or measured. The Python and R scripts that reproduce every figure and every fitted value of this work, together with the input tables and their provenance, are openly available at \url{https://github.com/robertmonjo/weak_lensing_flat_rotation}.

\smallskip
  \smallskip
 \smallskip

\appendix
\numberwithin{equation}{section}
\numberwithin{figure}{section}

\section{Minimally perturbed metric and circular orbital velocity}
\label{sec:circular}

This appendix derives the low acceleration limit from the weak-perturbed field given by the hyperconical model and examines the geodesic equations. Let $g$ be a first-order metric obtained from local GR assuming an approximately unitary scale factor ($t/t_0 \approx 1$) for gravitational systems with mass $M(r')$ enclosed in the radial coordinate $r'$ \citep{Monjo2024}:
\begin{eqnarray}
g \approx \left(1 - \int_{r'}^\infty \frac{2GM(r)}{c^2 r^2} dr\right) dt^2 - \left(1 + \int_{r'}^\infty \frac{2GM(r)}{c^2 r^2} dr\right) dr'^2 - r'^2 d\Sigma^2\,,
\end{eqnarray} where $\Sigma$ represents the angular coordinates with $d\Sigma = \sin\theta \, d\theta \, d\phi$. For instance, consider the most simple case with perturbation compatible with the $1/2$-distorting stereographic projection of \citet{MCS2023}, which is obtained from the following differential transformations assuming an almost constant $\gamma_0 = \gamma_0(r') \ge 2$:
\begin{eqnarray}
 r' &\to \hat r' \approx \left(1 + \frac{r'}{2\gamma_0(r') c t_0}\right) r' &  \quad  \quad  \Rightarrow  \quad  \quad   dr' \to d\hat r' \approx \left(1 + \frac{r'}{\gamma_0(r') c t_0}\right) dr'
\\
 t  &\to  \hat t \approx \left(1 + \frac{r'}{\gamma_0(r') c t_0}\right) t  &  \quad  \quad \Rightarrow   \quad \quad   dt \to  d\hat t \approx \left(1 + \frac{r'}{\gamma_0(r') c t_0}\right) dt + \,\text{neglected terms}
\end{eqnarray} Substituting these transformations into the metric $g = g_{ij}dx^i dx^j$, the resulting minimally-perturbed metric $\hat g$ is:
\begin{eqnarray} \label{eq:metric0}
\hat g &\approx& \left(1 - \int_{r'}^\infty \frac{2GM(r)}{c^2 r^2} dr + \frac{2r'}{\gamma_0 c t_0}\right) c^2 dt^2  - \left[\left(1 + \int_{r'}^\infty \frac{2GM(r)}{c^2 r^2} dr + \frac{2r'}{\gamma_0 c t_0}\right) dr'^2 + r'^2 \left(1 + \frac{r'}{\gamma_0 c t_0}\right) d\Sigma^2\right]\,.
\end{eqnarray} To analyse cosmic effects, we define an expanding coordinate $r \equiv (t/t_0) r'$, which is practically indistinguishable from the comoving coordinate $r'$ for gravitational systems, but it is a key for large scales. Therefore, the approaches $\frac{2GM}{c^2 r'} \approx \frac{2GM}{c^2 r}\ll 1$ and $\frac{2r'}{\gamma_0 c t_0} \approx \frac{2r}{\gamma_0 c t} \ll 1$ are assumed for the weak-field limit.


According to GR, the geodesic equations are expressed as $\ddot x^\mu + \Gamma^\mu_{\nu\sigma} \dot x^\nu \dot x^\sigma = 0$, where $\Gamma^\mu_{\nu\sigma} \equiv \frac{1}{2} g^{\mu\lambda} \left( \partial_\sigma g_{\lambda\nu} + \partial_\nu g_{\lambda\sigma} - \partial_\lambda g_{\nu\sigma} \right)
$ are the Christoffel symbols and $\dot x^\mu \equiv dx^\mu/d\tau$ is the derivative by the proper time $\tau$. The geodesic equations can also be derived from the Lagrangian $
\mathcal{L} = \left(g_{\mu\nu} \dot x^\mu \dot x^\nu\right)^{1/2}$ that provide the Euler-Lagrange equations $\frac{d}{d\tau} \left({\partial \mathcal{L}}/{\partial \dot{x}^\mu}\right) - {\partial \mathcal{L}}/{\partial x^\mu} = 0$. From this approach, a classical-like conservation of the energy $E$ is easily found as follows: \begin{eqnarray}
E \approx \left(1 -\int_{r}^\infty \frac{GM(r')}{c^2 r'^2} dr' + \frac{r}{\gamma_0 c t} - \frac{1}{2}\frac{\nu^2}{c^2}\right) m_0 c^2 \,.
\end{eqnarray}where $m_0$ is the mass of a test particle and $\nu$ is its speed. Since secondary terms of $c\,dt/d\tau \approx 1 + GM/(c^2r) -´r/(\gamma_0 c t) \approx 1$ only contribute to negligible terms in the geodesic equations, total acceleration of a test particle placed at the position $s = t e_t + r e_r$ with respect to a mostly central mass $M$ is\begin{eqnarray} \label{eq:accel}
   a_s \equiv  \frac{d^2 s}{c^2 dt^2}\approx  \frac{d^2 s}{d\tau^2} \approx - \frac{r}{\gamma_0 c t^2} e_t - \left( \frac{GM}{c^2 r} + \frac{1}{\gamma_0 c t}\right) e_r\,.
\end{eqnarray} Interpretation of Eq. \ref{eq:accel} is that centrifugal acceleration should also have both time-like and space-like components. Specifically, time-like component of the acceleration measured by an observer living in the hyperconical universe is expected to be proportional to $1/(ct)$ as the spatial contribution is proportional to $1/r$. Therefore, total centrifugal acceleration should be
\begin{eqnarray} \label{eq:centrif1}
a_s = \frac{d^2 s}{c^2dt^2} = -\frac{\omega_t^2}{c^2} ct \, e_t - \frac{\omega_r^2}{c^2}r\, e_{\vec r}   = -\frac{1}{ct} \frac{v_t^2}{c^2} \, e_t -\frac{1}{r}\frac{v_r^2}{c^2} \, e_{\vec r}  
\end{eqnarray} where $\omega_t^2 \equiv v_t^2/(ct)^2$ and $\omega_r^2 \equiv v_r^2/r^2$ are the squared angular speeds in the time- and space-like directions, respectively, while $v_t^2$ and $v_r^2$ are tangential squared speeds. Finally, we \textbf{hypothesize} that observers measure an effective centrifugal acceleration given by an apparent circular orbital speed $v$ around a radial distance $\mathcal{r}$, such as: \begin{eqnarray}
  \left| \left|a_s \right|\right| = \frac{1}{\mathcal{r}}\frac{v^2}{c^2} = - \frac{1}{\mathcal{r}} \left|\left|  \frac{v_t^2}{c^2}  \, e_t + \frac{v_r^2}{c^2} \, e_{\vec r} \right|\right|
\end{eqnarray}Thus, taking into account the Eqs. \ref{eq:accel} and \ref{eq:centrif1}, the absolute value of the speed $v$ is given by\begin{eqnarray} \nonumber
 \frac{v^4}{c^4} =
  - \left|\left|  \frac{v_t^2}{c^2}  \, e_t + \frac{v_r^2}{c^2} \, e_{\vec r} \right|\right|^2 & \approx &  -  \left[ \left( \frac{r}{\gamma_0  c t}\right)^2 - \left(\frac{\mathrm{G}M}{c^2r} + \frac{r}{\gamma_0 c t} \right)^2  \right] \approx \left(\frac{\mathrm{G}M}{rc^2}\right)^2 +\frac{2\mathrm{G}M}{\gamma_0 t c^3}
\;, 
\end{eqnarray}which satisfies two well-known limits of Newton's dynamics and the Milgrom's MOND:\begin{eqnarray}
\hspace{20mm} 
v \approx \sqrt{\frac{\mathrm{G}M}{ r}}\;\;  & \hspace{10mm} & \text{if}\; a_N(r) \gg \frac{2c}{\gamma_0 t} \simeq a_0 \\
\hspace{20mm} 
v \approx \sqrt[4]{\frac{2\mathrm{G}Mc}{\gamma_0 t}}\;\;  & \hspace{10mm} & \text{if}\; a_N(r)\ll \frac{2c}{\gamma_0 t} \simeq a_0 \,,
\end{eqnarray} where $a_N(r) \equiv GM/r^2$ is the classical Newtonian acceleration, $a_0$ is the Milogrom's constant acceleration and $M = M(r)$ is the total mass within the central sphere of radius $r$. Notice that the velocity curve $v=v(r)$ can be reworded in terms of the Kepler-Newton speed $v_N(r) := \sqrt{\mathrm{G}M(r)/r}$ leading to a mass-discrepancy acceleration relation:\begin{eqnarray} \label{eq:MOND}
v^4  = v_N^4 + v_N^2 \, r \frac{2c}{\gamma_0 t} \;\; \Longrightarrow\;\; \frac{M_\text{eff}(r)}{M(r)} = \frac{v^2(r)}{ v_N^2(r)} = \sqrt{1 + \frac{1}{|a_N(r)|} \frac{2c}{\gamma_0 t}}\;\;,
\end{eqnarray}with $M = r\,v_N^2/G$ is the baryonic mass and  $M_\text{eff} = r\,v^2/G$ is the apparent total or effective mass.

\section{Deflection angle in HMG} \label{sec:deflection_HMG}

This appendix summarizes the method of \citet{Gibbons2008} based on the Gauss-Bonnet theorem to derive the deflection angle ($\hat \varphi$) by integrating the Gaussian curvature ($K$) over a domain exterior to the lens, thus providing a geometrically insightful perspective on gravitational lensing by using its related optical metric. The Gauss-Bonnet theorem describes the geometry of a surface $D$ in relation to its topology, which is given by
\begin{eqnarray} \label{eq:theorem}
\iint_D K \, dA + \oint_{\partial D} \kappa_g \, ds + \sum_i \theta_i = 2\pi \chi(D) \quad \Rightarrow \quad \text{Weak lensing:}\; \hat \varphi  \equiv \sum_i \theta_i = - \iint_{D_2} K \, dA\,,
\end{eqnarray}where $k_g$ is the geodesic curvature, $\alpha_i$ are the exterior angles and $\chi(D)$ is the Euler characteristic for a domain $D$. For weak gravitational lensing, the Gaussian curvature $K$ is related to the gravitational potential around the lensing mass. The geodesic curvature $k_g$ is typically zero along the asymptotically flat boundary at infinity, contributing negligibly to the theorem. The exterior angles $\alpha_i$ account for the deficit angles due to spacetime curvature, directly relating to the light deflection angle $\hat  \varphi$, where $\sum_i \alpha_i = - \varphi$. The Euler characteristic $\chi(D)$ of the exterior domain is zero because the domain is non-compact and asymptotically resembles a flat plane with holes. By integrating the Gaussian curvature and accounting for the exterior angles, Gibbons and Werner’s method effectively captures how light bends around a gravitational lens, providing a straightforward and intuitive framework for calculating lensing effects in weak gravitational fields \citep{Zonghai2020}.


\begin{figure}[h]
\centering
	\includegraphics[width=0.75\columnwidth]{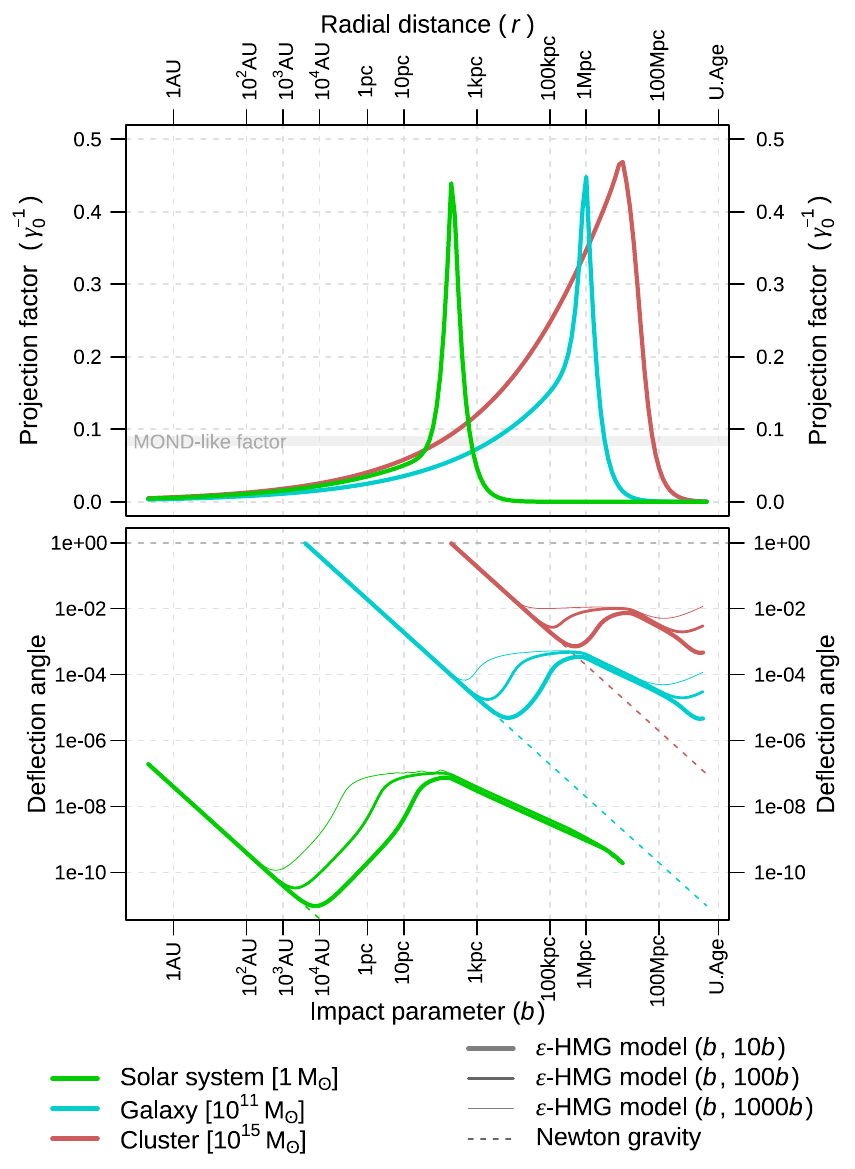}
    \caption{Example of projection factor (top panel) and corresponding deflection angle in radians (bottom panel) as a function of the impact factor ($b$) equal to the radial distance ($r$), for three gravitational systems centered, respectively, by three characteristic baryonic masses enclosed in $r$: Sun ($1 M_\odot$; green), galaxy ($10^{11} M_\odot$, blue) and galaxy cluster ($10^{15} M_\odot$, red). The models used are: the classical Newtonian gravity (dashed lines), and $\epsilon$-HMG (Eqs. \ref{eq:model_general2} and \ref{eq:sigma_model} with $r_\text{sys} \equiv r$ and different values of $r_\text{nei}/r_\text{sys}$ and $\zeta$; width colored lines). The represented $\epsilon$-HMG configurations ($r_\text{nei}/r_\text{sys}$, $\zeta$) are: (4, 0.9) for the Solar system, (3, 0.9) for the galaxy, and (2, 0.9) for the cluster. In the top panel, a MOND-like factor of $\gamma_0^{-1} \approx 1/12$ is shown in a thick gray band to represent the Milgromian constant $a_0 \approx 2\gamma_0^{-1}c/t \approx 1.2 \times 10^{-10}\, \text{m/s}^2$ The radial distances ($r$) range between 1 Astronomic Unit (AU) and the maximum \textit{light-travel distance} ($t_0c$) given by the age $t$ estimated for the Universe (noted as `U. Age', which is $t_0c \approx 13.7\,\text{Gyr}\,c$ with $c \equiv 1$). To solve the integral of Eq. \ref{eq:integral} in the bottom panel, the age of the Universe was considered with delay (i.e., $t \equiv t_0 - r/c$) and the domain of the radial distance ($r$) was ranged from $b$ to three different maximum values: $10 \times b$ (thickest colored lines), $10^2 \times b$ (intermediate colored lines) and $10^3 \times b$ (thinnest colored lines).}  \label{fig:FigA}
\end{figure}

In our case, let $g = g_{tt}dt^2 + g_{rr} dr^2$ be the metric defined by Eq. \ref{eq:metric0}. Both the $g_{tt}$ and $g_{rr}$ components contribute to the deflection angle of light. However, the time dilation represented by $g_{tt}$ plays a dominant role in determining the amount of bending that occurs, particularly because it directly represents the gravitational influence on the energy of photons. Finally, the interaction between the time dilation and spatial curvature effects (via the optical metric $\sigma$) leads to the observable phenomenon of gravitational lensing. Therefore, in weak-field approximations, where $2GM/c^2 \ll 1$, the Newtonian-like potential (from $g_{tt}$) gives the leading contribution to light bending, consistent with the classical expectation that gravity acts like a potential well that bending paths in its vicinity. Specifically, the optical metric $\sigma$ related to $g$ is defined by the null geodesics ($ds^2 = g_{ij}' dx^{i}dx^j = 0$), and simplifying for the equatorial plane ($\theta = \pi/2$) for outside the mass distribution, it is: \begin{eqnarray}
\sigma  =  \frac{g_{rr}}{g_{tt}} \, dr^2 + \frac{g_{\phi\phi}}{g_{tt}} \, d\phi^2 \;\approx \;\left(1 + \frac{4GM}{c^2 r} \right) dr^2 + r^2 \left(1 + \frac{2GM}{c^2 r}  - \frac{r}{\gamma_0(r)\,c\, t(r)} \right) d\phi^2\,,
\end{eqnarray} where Maclaurin expansion was used for the smallest terms and $t(r) = t_0 - r/c$ is considered to model effects from the early epochs of the Universe.  Assuming that $\partial_{r}\,\gamma_0(r) \ll \gamma_0(r)/r$, the Gaussian curvature $K$ is dominated by the crossing derivatives of $(r,\phi)$ over the component $\sigma_{\phi\phi} = g_{\phi\phi}/g_{tt}$ that are given by the Riemann tensor $R_{r'\theta r' \theta}$ of the optical metric $\sigma$, that is,
\begin{eqnarray}
 K = - \frac{R_{r\theta r \theta}}{\det \sigma} \approx \frac{2GM}{c^2 r^3} + \frac{1}{\gamma_0(r)\,r\,c\, t(r)} \,,
\end{eqnarray} where $\gamma_0(r) = \gamma_{sys}(r) / \cos \gamma_{sys}(r)$ and $\det \sigma = \sigma_{rr}\sigma_{\phi\phi} \approx 1$ is the determinant of the optical metric. The vaccum value of the projective factor is approximately $\gamma_0(r) \approx 2$ \citep{MCS2023}.

Since the area element is $dA = \sqrt{g_{rr}g_{\phi\phi}} dr\,d\phi \approx   r\,dr \, d\phi$, and integrating from $b/\sin(\phi)$ to the source of light (s.l), the deflection angle outside the mass distribution is:
\begin{eqnarray}
\label{eq:integral}
\hat \varphi &=& \int_{\delta}^{\pi-\delta} \int_{b/\sin(\phi)}^{\text{s.l.}} \left(\frac{2GM}{c^2 r^3} + \frac{1}{\gamma_0(r)\,r\,c\, t(r)}\right) r\, dr \, d\phi \;\; \approx \;\; \frac{4GM}{c^2 b} + \int_{\delta}^{\pi-\delta} \int_{b/\sin(\phi)}^{\text{s.l.}} \frac{\cos \gamma_\text{sys}(r)}{c\,t(r)\gamma_\text{sys}(r) } dr \, d\phi \,,
\end{eqnarray}
with $0 < \delta \ll 1$ and $t(r) = t_0 - r/c$. Considering Eq. \ref{eq:model_general2}, the contribution of the second part can be integrated by using standard numerical techniques (e.g., Simpson's rule), and it is shown in Fig. \ref{fig:FigA}.

For instance, the deflection angle for a typical galaxy ($M = 10^{11} M_\odot$ of baryonic mass) and impact parameter $b = 10 \, \text{kpc}$ is $\hat \varphi_{\text{GR}} \approx 10^{-5} \, \text{rad}$ according to GR. However, the contribution of HMG is $\hat \varphi_{\text{HMG}} > 10^{-4} \, \text{rad}$, which could be assimilated as dark-matter phenomenology at these distances. For the Sun ($M_\odot = 1.989 \times 10^{30} \, \text{kg}$) with an impact parameter of about $b = R_\odot = 6.96 \times 10^8 \, \text{m}$, the deflection angle is about $\hat \varphi_{\text{Sun}} \approx 1.75 \, \text{arcsec} \approx 8.5 \times 10^{-6} \, \text{rad}$, and $3.9 \times 10^{-8} \, \text{rad}$ for $b = 1$ AU. The cosmic contribution to the deflection angle predicted by the hyperconical model is about $10^{-15} \, \text{rad}$ for these short distances, and it is only significant from the radius of $10^3\,\text{AU}$ and extends effects beyond to $10^5\,\text{AU} \sim 0.5\,\text{pc}$, which corresponds approximately to the hypothetical Oort Cloud extension \citep{Crovisier2011}. Nevertheless, the predicted deflection angle for the outer solar system seems to be too large since it is extrapolated from the $\epsilon$-HMG parameters fitted with galaxies and clusters. Therefore, a specific analysis is required in future work.

\bibliography{00_main}{}
\bibliographystyle{aasjournal}

\end{document}